\documentclass[preprint,nofootinbib,12pt,floatfix,aps,prd,superscriptaddress]{revtex4-1}
\usepackage{amsmath,amssymb,amsfonts,amstext,amsbsy,amsopn}
\usepackage{graphicx, color, xcolor, slashed, esint}
\usepackage[dvips]{epsfig}
\usepackage{float, units, textcomp, hyperref}
\usepackage[utf8]{inputenc}
\usepackage{subcaption}
\usepackage[normalem]{ulem}
\usepackage{bm}
\newcommand{\beq}{\begin{equation}}
\newcommand{\eeq}{\end{equation}}
\newcommand{\bea}{\begin{eqnarray}}
\newcommand{\eea}{\end{eqnarray}}

\begin{document}

\title{Dymnikova Black Holes in Unimodular Gravity: Maxwell Sources and Vacuum Contributions}
\author{G. Alencar}
\email{geova@fisica.ufc.br}
\affiliation{Departamento de Física, Universidade Federal do Ceará, Caixa Postal 6030, Campus do Pici, 60455-760 Fortaleza, Ceará, Brazil}
\author{V. H. U. Borralho}
\email{victorborralho@fisica.ufc.br} 
\affiliation{Departamento de Física, Universidade Federal do Ceará, Caixa Postal 6030, Campus do Pici, 60455-760 Fortaleza, Ceará, Brazil}

\begin{abstract}
In this work, we investigate the Dymnikova regular black hole within the framework of unimodular gravity, emphasizing the role of the effective vacuum sector in the regularization of the geometry. By allowing a controlled violation of the covariant conservation of the energy--momentum tensor, the cosmological contribution emerges dynamically as a radial-dependent function, $\Lambda=\Lambda(r)$. We first reinterpret the Dymnikova spacetime as a charged configuration supported by nonlinear electrodynamics and derive the corresponding electric and magnetic sources. Subsequently, we demonstrate that the same geometry can be consistently generated by standard Maxwell electrodynamics in unimodular gravity. In this construction, the resulting electric field is everywhere regular and corresponds to a localized charge distribution with vanishing asymptotic charge, indicating that the spacetime does not behave as an asymptotically charged object. 
\end{abstract}
%
%\pacs{72.80.Le, 72.15.Nj, 11.30.Rd}
%
\maketitle
\section{Introduction}

The first exact solution of Einstein’s field equations was obtained in 1916 by Schwarzschild \cite{Schwarzschild1916}, describing a static and spherically symmetric spacetime and establishing the foundations of black hole physics. However, such geometries are generically characterized by spacetime singularities, where the classical description of gravity breaks down. In order to avoid these pathological features, several regular solutions have been proposed over the years \cite{Dymnikova:1992ux,Simpson:2018tsi,Hayward:2005gi}. The first regular black hole solution was proposed phenomenologically by Bardeen \cite{Bardeen1968}, describing a spacetime with a de Sitter core and asymptotically flat behavior. Later, Ayon-Beato and Garcia \cite{Ay_n_Beato_2000} showed that the Bardeen geometry can be generated by magnetic monopoles within nonlinear electrodynamics (NED). Other regular geometries, such as the Simpson--Visser and Dymnikova spacetimes \cite{Simpson:2018tsi,Dymnikova:1992ux}, were subsequently proposed. In parallel, considerable attention has been devoted to identifying the matter sources capable of supporting regular geometries \cite{Rodrigues:2023vtm,Alencar:2025jvl,Silva:2025fqj,Alencar:2024yvh,Fan:2016hvf,Fan:2016rih,Ay_n_Beato_2000,Rodrigues:2018bdc}. In this context, nonlinear electrodynamics has emerged as one of the most successful frameworks for obtaining regular black holes, including extensions to lower-dimensional black holes and black strings \cite{alencar2026blackstring,Bronnikov2023,Balart:2014cga,Cataldo:2000ns}.

Among the different regular geometries, the Dymnikova solution \cite{Dymnikova:1992ux} has attracted particular attention due to its simple realization of a regular black hole supported by an anisotropic matter distribution with a de Sitter core at the origin and Schwarzschild asymptotics at large distances. The corresponding energy--momentum tensor was phenomenologically introduced as
\begin{equation}
T_0^0 =T_1^1=\rho_0 e^{-r^3/r_*^3}, \qquad
T_2^2 =T_3^3=\rho_0 \left( 1- \frac{3r^3}{2r_*^3}\right)e^{-r^3/r_*^3},
\end{equation}
where $r_*^3=r_sr_0^2$, with $r_s$ and $r_0$ denoting the Schwarzschild and de Sitter radii, respectively. This construction replaces the Schwarzschild singularity by a regular de Sitter core while preserving asymptotic flatness. According to \cite{Dymnikova:1992ux,Ansoldi:2008jw}, the Dymnikova profile may be interpreted as a vacuum-induced configuration associated with gravitational vacuum polarization effects. In particular, the exponential behavior of the energy density was motivated by analogy with the Schwinger pair-production rate in quantum electrodynamics, $\Gamma \sim \exp(-E_c/E)$. More recently, the Dymnikova spacetime has been investigated in several contexts, including alternative matter sources \cite{Macedo:2025guc,Alshammari:2025hfm,Ahmed:2026vce}, effective quantum corrections and regular black-hole remnants \cite{Alencar:2023wyf,Konoplya:2023aph,Dymnikova:2015yma}, modified gravity scenarios \cite{Errehymy:2025djk,Konoplya:2024kih}, and gravitational perturbations \cite{Dubinsky:2025nxv}.

In another direction, unimodular gravity has attracted considerable attention as an alternative formulation of gravity in which the determinant of the metric is fixed, while the cosmological constant naturally emerges as an integration function of the field equations. Although classically equivalent to general relativity, the theory presents important conceptual and quantum differences that have been extensively discussed in the literature \cite{Bufalo:2015wda,Jirousek:2023gzr,Alvarez:2023eqo,Smolin:2009ti}. In particular, the unimodular constraint allows a modification of the standard covariant conservation law for the energy--momentum tensor, leading to an effective non-conservation equation explored in several cosmological and gravitational scenarios \cite{Alvarenga:2024yqa,Alvarenga:2025nwe,Piccirilli:2023klw,3137491,Alencar:2026oxy,Alencar:2026hjf,Fabris:2021atr}. Despite these appealing features, a first-principles motivation for unimodular gravity remained absent for a long time.

Recently, however, one of the present authors \cite{3122468} showed that the Einstein Equivalence Principle naturally implies a fixed spacetime volume element, $\sqrt{-g}=g_0$, identifying unimodular gravity as the framework most consistent with the geometric content of the equivalence principle. Within this perspective, several recent works have demonstrated that standard Maxwell sources can support regular black holes, black strings, and wormholes in unimodular gravity \cite{3137491,Alencar:2026hjf,Alencar:2026oxy}. Since Maxwell electrodynamics satisfies the standard energy conditions, the violations required for regularity can be entirely attributed to the vacuum sector. This is particularly relevant because, unlike the nonlinear electrodynamics sources commonly employed in regular black hole constructions, where the same matter sector both generates the geometry and violates the energy conditions, the unimodular framework allows a clean separation between the ordinary matter sector and an effective vacuum contribution encoded in the cosmological function.

However, up to now, the unimodular Maxwell framework described above has not been applied to the Dymnikova spacetime. In this work, we investigate whether the same construction can be consistently implemented for the Dymnikova black hole and explore its interpretation in terms of nonlinear electrodynamics and Maxwell sources within unimodular gravity. The paper is organized as follows. In Sec. II, we construct the nonlinear electrodynamics sources associated with the geometry. In Sec. III, we investigate its Maxwell realization within unimodular gravity and obtain the associated cosmological function. Finally, in Sec. IV, we present our conclusions and discuss possible perspectives.

\section{Nonlinear Electrodynamics Description of the Dymnikova Solution}
In this section, we reinterpret the Dymnikova geometry as a solution sourced by nonlinear electrodynamics (NED). The spacetime is described by the static and spherically symmetric metric
\begin{equation}\label{metric}
 ds^2=-f(r) dt^2+\frac{1}{f(r)}dr^2 +r^2d\theta^2 +r^2 \sin^2\theta d\phi^2,
\end{equation}
with
\begin{equation}
    f(r)=1-\frac{2M(r)}{r}, \qquad 
    M(r)=m\left(1-e^{-r^3/r_*^3}\right).
\end{equation}

To construct the corresponding matter sources, we consider the Einstein--Hilbert action coupled to a general nonlinear electrodynamics Lagrangian $L(F)$,
\begin{equation}
    S=\int d^4x \sqrt{-g}\Big[ R - L(F) \Big],
\end{equation}
where $R$ is the Ricci scalar and $F=F_{\mu\nu}F^{\mu\nu}$ is the electromagnetic invariant. The field strength tensor is described by the 2-form
\begin{equation}
    \bm{F} = \frac{1}{2} F_{\mu\nu} dx^\mu \wedge dx^\nu,
\end{equation}
which can be written in terms of the gauge potential $\bm{A}=A_\mu dx^\mu$ as $\bm{F}=\bm{dA}$. Varying the action with respect to the metric $g^{\mu\nu}$, we obtain the Einstein field equations:
\begin{equation}
    G_{\nu}^\mu =R_\nu^\mu-\frac{1}{2}\delta_\nu ^\mu R=T_{\nu}^\mu.
\end{equation}

For the spherically symmetric geometry in Eq.~\eqref{metric}, the nonvanishing components of the Einstein tensor are
\begin{align}
    G_0^0&=G_1^1=\frac{r f'(r)+f(r)-1}{r^2}, \\
    G_2^2&=G_3^3=\frac{r f''(r)+2 f'(r)}{2 r}.
\end{align}

The energy-momentum tensor is given by
\begin{equation} \label{TNED}
     T_{\mu}^{\ \nu}=
\delta_{\mu}^{\ \nu}\,\frac{L(F)}{2}
- 2\,L_{F}
\,F_{\mu}^{\ \lambda} F_{\lambda}^{\ \nu},
\end{equation}
where we define $L_F=dL/dF$.

Now, varying the action with respect to the gauge potential $A^\mu$, we obtain the generalized Maxwell equations:
\begin{equation}
    \nabla_\mu(L_FF^{\mu\nu})=0,
\end{equation}
which can be written in compact form using differential forms:
\begin{equation} \label{eqNED}
    \bm{d}(L_F \star \bm{F}) = 0, \quad \bm{d}\bm{F} = 0,
\end{equation}
where $\star$ is the Hodge operator. 

Thus, in order to obtain a NED source, we will interpret the Dymnikova solution as:
\begin{equation}
    f(r)=1-\frac{2m}{r} \left(1-e^{-r^3/Q^3}\right).
\end{equation}
where $Q^3=q^3+P^3$, in which $q$ represents the electric charge and $P$ represents the magnetic charge. 

In the following subsections, we will obtain the magnetic and electric sources.

\subsection{Purely magnetic source}

To obtain a purely magnetic solution, we must have $q=0$, $P\neq0$. In coordinates $(t,r,\theta,\phi)$, the ansatz for the 2-form is
\begin{equation}
    \bm{F} = P \sin\theta \ d\theta \wedge d\phi
\end{equation}
which trivially satisfies $\bm{dF}=0$. This configuration corresponds to a radial magnetic field, since the magnetic flux pierces spherical surfaces of constant radius, whose normal vector points along the $r$ direction. In this way, the electromagnetic invariant is written as
\begin{equation}
    F=F(r)=\frac{2P^2}{r^4},
\end{equation}
and in this case it is possible to invert this relation and obtain $r=r(F)$, simplifying the structure of the Lagrangian, which becomes $L=L(r).$

From the invariant $F(r)$, we can write the components of the energy-momentum tensor:
\begin{align}
     T_0^0 &=T_1^1= \frac{L(r)}{2}\\
     T_2^2 &=T_3^3= \frac{L(r)}{2}-2L_F\frac{P^2}{r^4}.
\end{align}
Thus, we obtain the equations:
\begin{equation} \label{EQ00}
    \frac{L(r)}{2}+\frac{6 m e^{-\frac{r^3}{P^3}}}{P^3}=0,
\end{equation}
\begin{equation}\label{EQ22}
   \frac{L(r)}{2}-\frac{2 P^2 L_F(r)}{r^4}+\frac{6 m e^{-\frac{r^3}{P^3}}}{P^3}-\frac{9 m r^3 e^{-\frac{r^3}{P^3}}}{P^6}=0.
\end{equation}

From Eqs.\eqref{EQ00} and \eqref{EQ22}, we obtain
\begin{align}
    L(r)=-\frac{12 m e^{-\frac{r^3}{P^3}}}{P^3} & \implies L(F)=-\frac{12m}{P^3}e^{-\left({8}/{P^6F^3}\right)^{1/4}} \\
    L_F(r)&=\frac{9 m r^7 e^{-\frac{r^3}{P^3}}}{2 P^8}
\end{align}

Taking the asymptotic limit, that is, $r\to 0 \;(F\to \infty)$, we obtain a constant value $L_0$:
\begin{equation}
    L(r) \approx-\frac{12 m}{P^3}+\mathcal{O}\left(r^3\right),
\end{equation}
following the proposition of Bronnikov \cite{Bronnikov2023}, according to which regular metrics require that $L(r) \to L_0$ in the limit $r\to0$. However, in linear limit, $r\to \infty \; (F\to0)$ we have $L(r) \approx0$.

\subsection{Purely electric source}
To obtain a purely electric source $(P=0, q\neq0)$, we consider the ansatz for the 2-form
\begin{equation}
    \bm{F}=E(r)\, dt \wedge dr,
\end{equation}
where $E(r)$ represents the radial electric field in a static and spherically symmetric configuration. This choice automatically ensures the absence of magnetic components, reducing the nonlinear electrodynamics sector to its purely electric sector. In this case, the electromagnetic invariant is given by
\begin{equation}
    F(r)=-E^2(r),
\end{equation}
which depends only on the radial coordinate due to the symmetry of the system.

Solving the generalized Maxwell equations in the presence of a nonlinear Lagrangian $L(F)$, we obtain the following relation
\begin{equation}
   \bm{d}\!\left(L_{F} \star \bm{F}\right)
    =
    \partial_r\!\left[r^2 \sin^2\theta \;L_{F} E(r) \right]
    dr \wedge d\phi \wedge dz
    =
    0 .
\end{equation}
The structure of this equation reflects the conservation of the nonlinear electric displacement field. Since $dr \wedge d\phi \wedge dz \neq 0$, the radial equation reduces to an ordinary differential constraint, yielding
\begin{equation}\label{campoeletrico}
    E(r)=\frac{q}{r^2 L_{F}},
\end{equation}
where $q$ is an integration constant associated with the electric charge of the configuration. Here, unlike the purely linear Maxwell case, the nonlinear nature of the theory prevents us from expressing $r$ explicitly as a function of $F$, since this inversion depends on the specific form of $L_{F}(F)$.

Thus, we can write the components of the energy-momentum tensor as
\begin{align}
    T_0^0&=T_1^1=\frac{L(r)}{2}+\frac{2q^2}{r^4 L_{F}}, \\
    T_2^2&=T_3^3= \frac{L(r)}{2},
\end{align}
where the anisotropy between radial and angular pressures is a direct consequence of the presence of the electric field.

Then, the Einstein field equations for the chosen static and spherically symmetric geometry become:
\begin{equation} \label{eletricEQ00}
 \frac{L(r)}{2}+  \frac{2q^2}{r^4 L_{F}}+\frac{6 M e^{-\frac{r^3}{q^3}}}{q^3}=0,
\end{equation}
\begin{equation} \label{electricEQ22}
    \frac{L(r)}{2}+\frac{6 m e^{-\frac{r^3}{q^3}}}{q^3}-\frac{9 m r^3 e^{-\frac{r^3}{q^3}}}{q^6}=0.
\end{equation}

From the above equations, we obtain the purely electric source
\begin{equation}
    L(r)=-\frac{6 m e^{-\frac{r^3}{q^3}} \left(2 q^3-3 r^3\right)}{q^6} \quad , \quad L_F(r)=-\frac{2 q^8 e^{\frac{r^3}{q^3}}}{9 m r^7}.
\end{equation}
 Near the origin and in the asymptotic limit, the behavior coincides with that of the magnetic case: $L(r\to0)\approx-12m/q^3 +\mathcal{O}(r^3)$ and $L(r \to\infty)\approx0$.
Naturally, after obtaining the Lagrangian, we can write the electric field from Eq.~\eqref{campoeletrico}
\begin{equation}
    E(r)=-\frac{9 m r^5 e^{-\frac{r^3}{q^3}}}{2 q^7}.
\end{equation}
In the limit $r\to0$, $E(r)\approx0$, demonstrating the regularity of the solution. Moreover, in the asymptotic limit, $E(r\to\infty)\approx0$, in agreement with the theorem proposed by Bronnikov stating that a NED Lagrangian describing a spherically symmetric spacetime regular at the origin cannot exhibit the Maxwell limit as $r\to\infty$ \cite{Bronnikov2023}. From these results, we see that the electric field corresponds to a localized distribution, with its peak located at $r_c=(5/3)^{1/3}q$, such that $E(r_c)=-{9 m e^{-5/3}\left({5}/{3}\right)^{5/3}}/{2 q^2}$.

\begin{figure}[H]
    \centering
    \includegraphics[width=0.5\linewidth]{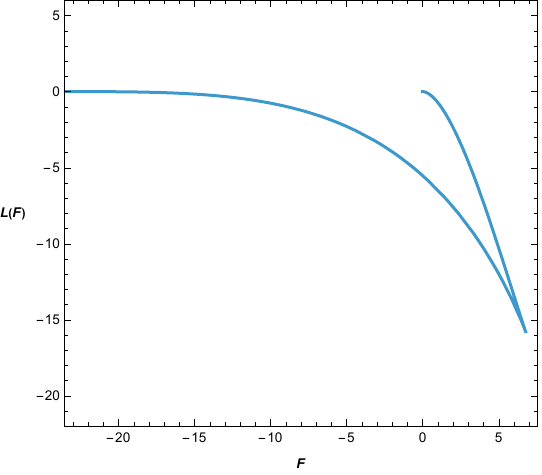}
    \caption{Behavior of $L(F)$ as a function of the scalar $F$. We use $m=2$ and $q=1$.}
    \label{lag1}
\end{figure}

It is important to note that the electric configuration does not lead to a unique Lagrangian $L(F)$. In fact, the relation $F=F(r)$ is not globally invertible, since the same value of $F$ can correspond to different radial positions. Consequently, the substitution $r=r(F)$ leads to different branches of $L(F)$, rather than to a single-valued function. Therefore, the electric NED description of the Dymnikova geometry cannot be represented by a unique Lagrangian $L(F)$ over the entire spacetime. See Fig.\eqref{lag1}.

\section{Maxwell Source: unimodular gravity Extension}
In this section, we will obtain a Maxwell source for the Dymnikova black hole by using the unimodular gravity framework. Unimodular gravity  allows one to relax the conditions on the energy-momentum tensor; in this way, by imposing its non-conservation, it is possible to obtain an effective electromagnetic source associated with the cosmological term \cite{3137491}.

In four dimensions, the field equations in unimodular gravity are written as \cite{Smolin:2009ti}:
\begin{equation}
    R_{\mu\nu}-\frac{R}{4}g_{\mu\nu}=T_{\mu\nu}-\frac{1}{4}g_{\mu\nu}T. 
\end{equation}
where $T=T_\mu^\mu$ is the trace of the energy-momentum tensor. Taking the divergence of both sides of the expression, together with the condition $\nabla_\mu T^{\mu\nu}=0$, we obtain
\begin{equation}
   \partial_{\nu}(R+T)=0 \implies R+T=const.,
\end{equation}
so that the integration constant can be defined as $-4\Lambda$, and we recover the Einstein field equations of general relativity with a cosmological constant:
\begin{equation} \label{EFE}
    G_{\mu\nu}=T_{\mu\nu}+\Lambda g_{\mu\nu}.
\end{equation}

As mentioned earlier, unimodular gravity allows one to relax the conditions on the energy-momentum tensor, and in this way, we can write the Maxwell equations with
\begin{equation}
    \nabla_\mu F^{\mu\nu}=J^\nu,
\end{equation}
which correspond to Maxwell's equations in the presence of sources, where $J^\nu$ is the four-current describing the distribution and flow of electric charge. Taking the divergence of Eq.~\eqref{EFE}, we obtain
\begin{equation}
    \nabla_\mu T^{\mu\nu} = -F^{\nu\alpha} J_\alpha.
\end{equation}
In this way, the cosmological term becomes dynamical, $R+T=-4\Lambda(x)$, and the equations become more general:
\begin{equation}
     G_{\mu\nu}=T_{\mu\nu}+\Lambda (x) g_{\mu\nu}.
\end{equation}
such that $T^{\text{eff}}_{\mu\nu}=T_{\mu\nu}+\Lambda (x) g_{\mu\nu}$. Thus, since the components of the energy-momentum tensor for Maxwell electrodynamics are already known, once we determine the function $\Lambda(x)$, we will have all the information about the matter content that sources the Dymnikova solution.

In addition, we can extract the relation
\begin{equation}
    \partial_\nu \Lambda = F_{\nu\alpha} J^\alpha,
\end{equation}
showing that the cosmological term behaves as an effective electromagnetic source (we will consider $\Lambda = \Lambda(r)$ due to the symmetry of the problem). By considering a purely magnetic configuration characterized by $F_{\theta\phi}=P\sin\theta$, the corresponding radial contribution to the cosmological function vanishes, since $F_{r\alpha}J^\alpha=0$. Therefore, within the Maxwell sector of unimodular gravity, the magnetic field considered here does not induce a radial variation of $\Lambda$, yielding $\partial_r\Lambda=0$ \cite{3137491}.

Before obtaining the function $\Lambda(r)$, it is necessary to verify the validity of the solutions in the Maxwell regime. For this purpose, in Ref.~\cite{3137491}, starting from Einstein's equations, the geometric function
\begin{equation}\label{Hfunction}
    H(r) \equiv E^2(r) = \frac{M'(r)}{r^2} - \frac{1}{2}\frac{M''(r)}{r}
\end{equation}
is defined, such that, in order to ensure that the electric field is always real, the condition for the validity of the sources in the Maxwell regime is $H(r) \geq 0$. In the Maxwell regime, the energy-momentum tensor is traceless, $T=0$, and therefore
\begin{equation}
    \Lambda(r)=\frac{R}{4}.
\end{equation}
This result is particularly significant: in the Maxwell regime, the vacuum contribution, namely the cosmological term, is not an independent degree of freedom, but instead emerges entirely from the spacetime geometry, being completely determined by the Ricci scalar.
Now, computing the function $H(r)$, we obtain
\begin{equation}
  H(r)=\frac{9 m r^3 e^{-\frac{r^3}{q^3}}}{2 q^6},
\end{equation}
so that $H(r)>0$ for all $r$, confirming the validity of the electric source in the Maxwell regime. This result ensures that the electric field remains real everywhere, which is a necessary condition for the physical consistency of the solution.

In the unimodular gravity framework, the electric field associated with the Dymnikova geometry acquires a particularly interesting interpretation. Starting from Eq.\eqref{Hfunction}, we obtain the electric field:
\begin{equation}
E(r)=\pm \frac{3\sqrt{m/2}}{q^3}r^{3/2}e^{-r^3/(2q^3)}
\end{equation}
is everywhere regular, vanishing both at the origin and asymptotically, and therefore does not exhibit the usual Coulomb behavior. Moreover, since the effective charge function is given by
\begin{equation}
Q(r)=r^2E(r),
\end{equation}
the total electric charge measured at spatial infinity satisfies
\begin{equation}
Q_\infty=\lim_{r\to\infty}Q(r)=0.
\end{equation}
Thus, the electromagnetic sector describes a localized charge distribution with exponentially suppressed flux at large distances, rather than an asymptotically charged object. In this construction, the Maxwell field itself remains physically regular, while the regularization of the spacetime is entirely encoded in the effective vacuum contribution described by the radial-dependent cosmological function $\Lambda(r)$. 

Finally, computing $\Lambda(r)$, we find
\begin{equation}
    \Lambda(r)=\frac{3 m e^{-\frac{r^3}{q^3}} \left(4 q^3-3 r^3\right)}{2 q^6}.
\end{equation}
Near the origin, $\Lambda(r)$ approaches a constant value,
\begin{equation}
    \Lambda(r) \approx \frac{6 m}{q^3},
\end{equation}
indicating the presence of an effective de Sitter core, as already known for the Dymnikova black hole. This behavior is directly related to the regularity of the geometry at $r=0$, preventing the formation of curvature singularities.

On the other hand, in the asymptotic limit $r \to \infty$, the exponential suppression leads to
\begin{equation}
    \Lambda(r) \to 0,
\end{equation}
which shows that the spacetime becomes asymptotically flat, with a vanishing effective cosmological term.

Therefore, the cosmological function $\Lambda(r)$ smoothly interpolates between a constant value near the origin and zero at infinity, revealing that the effective vacuum contribution is entirely determined by the spacetime geometry. In particular, this result reinforces the interpretation that, in the Maxwell regime of unimodular gravity, the cosmological term is not fundamental, but rather an emergent geometric quantity.

\section{Conclusions}

In this work, we investigated the Dymnikova black hole within the framework of unimodular gravity. As a first step, we reinterpreted the parameter $r_*^3$ as an effective charge scale and constructed the corresponding nonlinear electrodynamics sources associated with the geometry. In particular, we obtained both purely magnetic and purely electric configurations, while no dyonic source can generate the Dymnikova spacetime, consistently with Bronnikov’s no-go theorem for regular black holes sourced by nonlinear electrodynamics \cite{Bronnikov2023}. This construction then allowed us to investigate whether the same geometry could be generated by standard Maxwell electrodynamics together with an effective vacuum contribution within unimodular gravity.

Within the unimodular framework, we then showed that the same geometry can be generated by standard Maxwell electrodynamics through an effective non-conservation of the energy--momentum tensor. This mechanism gives rise to a radial-dependent cosmological function $\Lambda(r)$, which behaves as an effective vacuum contribution. Near the origin, the function reproduces the de Sitter core responsible for the regularity of the geometry, while asymptotically it vanishes, recovering the expected Schwarzschild behavior. 

A particularly relevant aspect of this construction is that, unlike the usual nonlinear electrodynamics descriptions of regular black holes, the unimodular Maxwell realization allows a clean separation between the ordinary matter sector and the vacuum contribution responsible for regularization. Since Maxwell electrodynamics satisfies the standard energy conditions, the violations required for the existence of a regular core are entirely encoded in the effective vacuum sector described by $\Lambda(r)$. This result reinforces previous findings obtained by some of the present authors in the context of regular black holes, black strings, and wormholes in unimodular gravity \cite{3137491,Alencar:2026hjf,Alencar:2026oxy}, where the same mechanism was shown to occur. In this sense, the present construction provides a natural realization of the original Dymnikova interpretation, where the regular geometry is associated with vacuum effects, while explicitly separating the ordinary Maxwell source from the vacuum contribution that emerges dynamically through unimodular gravity.

\acknowledgments{The author G.A. would like to thank Conselho Nacional de Desenvolvimento Cient\'{i}fico e Tecnol\'{o}gico (CNPq) and Fundação Cearense de Apoio ao Desenvolvimento Científico e Tecnológico (FUNCAP). V.H.U.B. is supported by Coordena\c c\~{a}o de Aperfei\c coamento de Pessoal de N\'{i}vel Superior - Brasil (CAPES) - Financial Code - 001.
}
\bibliographystyle{apsrev4-1}
\bibliography{referencias}
\end{document}